# DEPENDENCE OF RESONANT ABSORPTION LINEWIDTH ON ATOMIC VAPOUR COLUMN THICKNESS FOR $D_1$ LINE OF Cs ATOMS CONFINED IN NANO-CELL

A. Sargsyan[1], D. Sarkisyan[1], Y. Pashayan-Leroy[2], C. Leroy[2], P. Moroshkin[3], A. Weis[3]

[1] Institute for Physical Research, NAS of Armenia-0203, Ashtarak, Armenia;
[2] Institut Carnot de Bourgogne - UMR CNRS 5209, Universite de Bourgogne 47870, F-21078 Dijon Cedex, France
[3] Department of Physics, University of Fribourg, 1700 Fribourg, Switzerland

A new nano-cell with smoothly varying longitudinal thickness of the atomic vapour layer $L$ in the range of 350 - 5100 nm allowing to study the resonant absorption of $D_1$ and $D_2$ lines of Cs atoms for thicknesses changing from $L = \lambda/2$ to $L = 6\,\lambda$ with the step of $\lambda/2$ ($\lambda = 852$ nm or 894 nm are the resonant laser wavelengths) and for different intensities is developed. It is revealed that for low laser intensities there is narrowing of the resonant absorption spectrum for the thicknesses $L = (2n + 1)\,\lambda/2$ (where $n$ is an integer) up to $L = 7\,\lambda/2$ and broadening of the spectrum for $L = n\lambda$. For relatively high laser intensity (>1 mW/cm$^2$), velocity selective optical pumping/saturated resonances of a reduced absorption (with the line-width close to the natural one), and centred on the hyperfine transitions occur when $L = n\,\lambda$. The possible application of these resonance peaks is given. The developed theoretical model describes well the experiment.

PACS numbers: 42.50.Gy, 42.62.Fi, 42.50.Md

## 1. Introduction

For the last years there have appeared a number of articles dedicated to the laser spectroscopy (processes of resonant absorption and fluorescence) of atomic vapor column of 10÷100 µm [1], and of atomic vapor column contained in so called extremely thin cells or nano-cells [2-7]. It was shown that an important parameter determining the spectral line-width, line-shape and the magnitude of absorption in such cells is the parameter $L / \lambda$, where $L$ is the thickness of the atomic vapor column (i.e. the distance between the windows of the nano-cell) and is the laser wavelength resonant with the atomic transition. In particular, it was shown that the spectral line-width of the resonant absorption is minimal for $L = (2n +1)\,\lambda/2$ (where n is integer), which was called Dicke-type Coherent Narrowing Effect (DCNE). Also, it was shown [3-7] that for $L = n\,\lambda$ the spectral line-width of the resonant absorption reaches the maximal value close to the Doppler width (about several hundreds of MHz) registrated in ordinary cells (1 – 10 cm long) [8]. This effect was called the collapse of DCNE. In [5] the DCNE effect and the collapse of DCNE were investigated up to the thicknesses $L = 7\lambda/2$ for $D_2$ line of $^{85}$Rb, $\lambda = 780$ nm. We have developed a new nano-cell allowing to change smoothly the thickness of the atomic vapor column in the range of 350÷5000 nm which allows one to study the greater values of $L / \lambda$. Note that earlier in [2-5] the resonant absorption of D2 lines of Rb и Cs atoms for which the frequency distance between the upper hyperfine levels is rather small (about tens of MHz) that made technically difficult the study of individual atomic transitions was investigated. From this point of view more convenient are the D1 lines of Rb и Cs for which the frequency distance between upper hyperfine levels reaches ~1 GHz that makes the atomic transitions completely resolved.

## 2. Experimental setup

The design of the nano-cell with a wedge-shaped (variable) gap between the windows is similar to the one presented in [2-5], however, the thickness of the atomic column could be changed in the range of 350÷5000 nm that is much more than that in [2-5]. The increase of the overlapped range of thicknesses was achieved by a preliminary evaporation of $Al_2O_3$ layer with the thickness of ~ 5000 nm on the surface of one of the windows of the nano-cell in its lower part. The wedge-shaped (along the vertical) thickness of the gap of the nano-cell was determined by the interferometric method described in [3]. The scheme of the experimental setup is presented in Fig. 1. The beam of a frequency-tunable diode laser (Distributed Feedback - DFB) with the wavelength $\lambda = 894$ nm, the spectral width $\gamma_L \sim 6$ MHz and the diameter of 1 mm irradiated the nano-cell under the angle close to the normal. The laser intensity was regulated with neutral filters F (as shown in Fig. 1). The nano-cell was inserted into a complementary oven having two apertures for laser passage. The temperature of the side- arm of the nano-cell determining the density of the Cs atoms was 110 ÷120 0C. To prevent atomic vapour condensation on the cell windows the temperature of the windows was hold higher by 20÷30 degrees. The laser frequency was scanned in the range ~10 GHz, overlapping all four spectral components: Fg = 3 → Fe = 3, 4 и Fg = 4 → Fe = 3, 4 of the hyperfine structure of $D_1$ line of Cs (Fg, Fe denote lower and upper D1 lines, respectively). To obtain transmission spectra for different thicknesses $L$ the oven with the cell was shifted along the vertical as shown by the arrow in Fig. 1. The small part of the radiation was sent to the cell of ordinary length (~ 3 cm), hold at the room temperature, and in which the known scheme of Saturation Absorption (SA) was realized [8]. The filters F were used to weaken the probe laser (also the pump laser) in the SA scheme in order to get the spectral reference with the width close to the natural one. The spectra were registrated with two-beam digital oscilloscope Tektronix.

## 3. Theoretical model

Consider the interaction of an atomic gas with a linearly polarized electric field of the laser radiation propagating in the $z$ direction $E(z) = E_0 \cos(\omega_L t - kz)$ with $\omega_L$ being the frequency, $E_0$ the amplitude and $k$ the wave number. The atomic gas is confined in an extremely thin cell with the thickness $L$. The equation describing the evolution of the density matrix in the external field $E(z)$ is

$$\frac{\partial \rho}{\partial t} = -\frac{i}{\hbar}\left[\bar{H}_0 + \bar{V}, \rho\right] + \hat{\Gamma}\rho, \tag{1}$$

where $\bar{H}_0$ the Hamiltonian of a free atom. For the interaction with the field $E$ we restrict ourselves by electric dipole approximation $V = -dE$ ($d$ is the atomic dipole moment). Here $\hat{\Gamma}$ describes the relaxation processes. We consider the atoms to be two-levels of definite (opposite) parity ($d_{11} = d_{22} = 0$). In the rotating-wave approximation Eq. (1) reads

$$\frac{\partial \rho_{11}}{\partial t} = -2\,\mathrm{Im}(\Omega^* \rho_{21}) + \gamma_{21}\rho_{22},$$

$$\frac{\partial \rho_{22}}{\partial t} = 2\,\mathrm{Im}(\Omega^* \rho_{21}) + \Gamma_{22}\rho_{22}, \tag{2}$$

$$\frac{\partial \rho_{21}}{\partial t} = i\Omega(\rho_{11} - \rho_{22}) - (i\Delta + \Gamma_{21})\rho_{21},$$

where $\gamma_{21}$ is the rate of spontaneous decay from level 2 to level 1, $\Gamma_{22} = \gamma_{21} + \gamma_0$ is the total decay rate of level 2, $\gamma_0$ is the rate of population lost from the system leading to the optical pumping of the other fundamental sublevel $6\,^2S_{1/2}$, $\Gamma_{21} = \frac{1}{2}(\gamma_{21} + \gamma_0)$ is the total relaxation rate of the coherence $\rho_{21}$. Here $\Delta = (\omega - \omega_{21} - \mathbf{kv})$ is the frequency detuning with the Doppler shift $\mathbf{kv}$ taken into account, with $\mathbf{v}$ being the atomic velocity, and $\Omega = -E_0 d / \hbar$ the Rabi frequency. To take into account the laser bandwidth we use the phase diffusion model of Wigner-Levy [9-11], in accordance with which it is assumed that the laser radiation has a Lorentzian spectrum with the Full Width at Half Height (FWHH) of $\gamma_L$. The bandwidth is incorporated into Eqs. (2) as a relaxation term for the non-diagonal element of the density matrix in accordance to the procedure given in [12,13]. The absorption spectrum is calculated by numerical integration of Eqs. (2) for the density matrix with subsequent averaging over the ensemble of atomic velocities that is supposed to be Maxwellian. The theoretical model and the formula used are presented in [6,7]. The main assumptions made in the model are as follows: the atomic number density is assumed to be low enough so that the effect of collisions between the atoms can be ignored; the atoms experience inelastic collisions with the cell walls, i.e. atoms lose completely their optical excitation; the incident beam diameter largely exceeds the cell thickness which allows one to neglect the relaxation of atoms traveling out of the diameter of the laser beam. The effect of the reflection of the radiation from the highly-parallel windows of the nano-cell behaving as a Fabri-Perot cavity [6] is taken into account.

## 4. Results and discussions

Fig. 2 presents the experimental transmission spectra of the resonant absorption of the nano-cell for the atomic transition Fg = 4 → Fe = 3 at the laser effective intensity (EI) $I_{\text{eff}} \sim 0.03$ mW/cm$^2$ for the cell thicknesses $L = \lambda/2$ and $L = 3$ cm. EI is determined by multiplying the measured intensity by the coefficient $[\gamma_N/(\gamma_N + \gamma_L)]$, where $\gamma_N$ is the spontaneous decay of Cs D$_1$ line and $\gamma_L \sim 6$ MHz . Curve 1 is the transmission spectrum of the nano-cell for the arm-side temperature of $\sim 110 \div 120\,^0$C. The absolute magnitude of the absorption is 1-2%. The experimental profile of transmission curve *1* is best approximated by curve *2* which is described by "pseudoVoight2" function of ORIGIN 7, whereby the Lorentzian and Gaussian profiles have the Full Width at Half Maximum (FWHM) of 70 MHz and 77 MHz , respectively (meanwhile, the parameters m and A of the "pseudoVoigt2" function were also fitted). Curve *2* has the FWHM ~75 MHz. Note, that the theoretical curve in Fig. 3b at $L = \lambda/2$ is also well approximated by the "pseudoVoigt2" function, which is an additional evidence of that the chosen theoretical model is correct. Curve 3 is the experimental Doppler broadened transmission spectrum of the nano-cell with $L = 3$ cm, which is well described by Gaussian profile *4* with FWHM ~350 MHz (in the figure curves *3* and *4* are practically merged). Hence, at $L = \lambda/2$ there is narrowing of the spectrum by a factor of 4.6. If one compares the transmission spectra of the nano-cell with those of the ordinary cell at the same temperature of the windows ~ 150 $^0$C, the FWHM for the ordinary cell is 430 MHz and the spectrum will be narrowed by a factor 5.7. Fig. 3a shows the experimental transmission spectra for the same atomic transition when changing $L$ from $\lambda/2$ to $5\lambda$ with the step of $\lambda/2$ at EI of ~ 0.03 mW/cm$^2$. The effect of DCNE and collapse of DCNE is well seen in the figure: the minimal value of the spectral width is achieved for the column thickness $L = (2n +1) \lambda/2$. Note that at $L = \lambda/2$ the spectral width is minimal and is about ~ 75-80 MHz [3-7], meanwhile at $L = 3\lambda/2$ , $5\lambda/2$, $7\lambda/2$ and $9\lambda/2$ spectral narrowing occurs in the presence of increasing wide Doppler pedestal (obviously, at further increase of $L$ the spectrum will more and more resemble Doppler broadened absorption spectrum of the ordinary cell). As seen from the figure at $L = n\lambda$ the spectral width is maximal (at $L = \lambda$ the spectral width is ~300 MHz). Hence, there occurs 4-multiple narrowing of the spectrum at $L = \lambda/2$ as compared with that of $L = \lambda$. Note that in [3] for $D_1$ line there was a 2-multiple narrowing of the spectrum which apparently is explained by the following: because of small gaps between the nano-cell windows the pumping out of the residual gas (absorbed by the windows surface) becomes difficult and that is why a many-hours pumping out is required at the temperatures of the windows of about ~ 400 $^0$C; in [3] the first model of the nano-cell was used in which the pumping out of the residual gas was probably worse than in the subsequent ones. The residual gases can lead to an additional broadening of the transmission spectrum and this affects more at the thickness $L = \lambda/2$ than at $L = \lambda$ . Note that the peak transmission at the exact resonance for $L = \lambda/2$ is practically equal to the peak transmission $L = \lambda$, meanwhile with increase of the laser intensity (see Figs. 4, 5) the peak transmission at $L = \lambda/2$ becomes more that of $L = \lambda$. The lower curves of Figs. 3a, 4a, 5a were obtained by the known Saturation Absorption method in the cell of ordinary length. Fig. 3b shows the theoretical spectra of the resonant transmission for the same transition at the Rabi frequency $\Omega = 0.01$ $\gamma_N$; $\gamma_N \sim 4.56$ MHz , $\gamma_L \sim 6$ MHz , and the thermal velocity $v_T = 200$ m/s. As it is seen there is good agreement between the experiment and the theory.

Fig. 4a presents the experimental spectra of the resonant transmission with all the parameters the same as in Fig. 3a but for EI of the laser ~ 0.5 mW/cm². The effect of CDNE and the collapse of CDNE is well seen in the figure: the minimal spectral width is achieved at the atomic vapor column $L = \lambda/2$, $3\lambda/2$ , and $5\lambda/2$. However, for $L = n\lambda$ in distinction from Fig. 3 at exact atomic resonance there appears so-called VSOP (Velocity Selective Optical Pumping/Saturation) peak. This peak of decreased absorption is located exactly at the atomic transition [1, 4, 5], and arises because the atom in the ground level Fg = 4 absorbs a laser photon and goes to the excited level Fe = 3, and further goes spontaneously to the ground level  Fg = 3 or to Fg = 4. This effect is called optical pumping (OP) [7, 14].  As a result of OP the part of the atoms goes to level Fg = 3 and the number of atoms which absorb from level Fg = 4 decreases, which leads to the decrease of the absorption from this level. The efficiency of OP is determined by the expression [7]

$$\eta \sim \frac{\Omega^2 \gamma_N t}{\left(\Delta + \mathbf{k}\mathbf{v}\right)^2 + \Gamma^2} \tag{3}$$

where $t$  is the time of interaction of the radiation with the atom, $\Delta$  is the detuning of the resonance, $\Gamma$ is the sum of homogeneous and non-homogeneous broadenings. From (3) it is seen that more the interaction time $t$, more the efficiency of OP. For the atoms flying perpendicularly to the laser beam the interaction time $t_D = D / v$ ($D$ is the laser beam diameter), meanwhile for the atoms flying along the laser beam the interaction time is $t_L = L / v$.  As the laser beam diameter is ~1 mm, and the distance between the walls $L = 894$ nm, $t_D$ is more than $t_L$ by three ordes.  For the atoms flying perpendicularly to the laser beam $\mathbf{k}\mathbf{v} = 0$, and expression (3) takes the maximal value at $\Delta = 0$. For this reason VSOP peak is located exactly at the atomic transition [1, 4, 5]. Fig. 4b presents the theoretical spectra for the same transition for the Rabi frequency $\Omega = 0.06\ \gamma_N$, the other parameters are the same as in Fig. 3b. As seen there is a good agreement between the theory and the experiment. Fig. 5a shows the experimental transmission spectra with all the parameters the same as in Fig. 3a, but EI of the laser is ~ 7 mW/cm$^2$. The effect of CDNE is only observed for $L = \lambda/2$, and for the other values of $L$ there appear additional VSOP peaks because as seen from formula (3) the increase of the laser intensity leads to the increase of the efficiency of OP. In Fig. 5b we present the theoretical spectra for the same transition at $\Omega = 0.2\ \gamma_N$, the other parameters are the same as in Fig. 3b. There is also good agreement between the theory and the experiment. It is important to note that at low laser intensities the spectral width of the VSOP resonance is close to the natural width. It is well seen from the comparison of the spectral widths of VSOP resonance at $L = \lambda$  with the peak of decreased absorption obtained by SA technique (Fig. 3a). That is why from the practical point of view the VSOP resonances (at $L = \lambda$ the minimal spectral width is achieved) could be used as the reference marks (with unidirectional laser beam) for atomic transitions as it is realized with the help of SA method. Besides, in external magnetic fields VSOP resonance can be spitted in several new peaks and the peaks' amplitudes and their frequency position depend on the magnitude of the applied external field [15, 16]. This will allow one to study the atomic transitions between Zeeman sublevels in the transmission spectra of Cs $D_1$ lines.

The present work has been submitted to ArXiv when a new paper concerning the peculiarities of the absorption and the fluorescence in a nanocell with thickness $L = \lambda/2$ and $L = \lambda$ has been published [17].

A. S. is grateful to the Institute of Physics of the University of Fribourg, Switzerland for the possibility to perform experimental study. A.S, D.S, P. M and A.W are thankful for the financial support provided by SCOPES Grant IB7320-110684/1. The authors A.S, D.S, Y. P-L, и C.L are thankful for the financial support provided by INTAS South-Caucasus Grant 06-1000017-9001.

References

1. S. Briaudeau, D. Bloch, M. Ducloy, Phys. Rev. A 59, 3723 (1999)

2. D.Sarkisyan, D.Bloch, A.Papoyan, M.Ducloy, Opt. Commun., 200, 201 (2001).

3. 3 .D. G. Dutier, A. Yarovitski et al., Europhysics Letters, 63 (1), 35 (2003).

4. Sarkisyan, T.Varzhapetyan, at al., Phys. Rev. A 69, 065802 (2004)

5. Sarkisyan, T. Varzhapetyan, at al., Proc. SPIE 6257, 625701 (2006).

6. G. Dutier, S. Saltiel, D. Bloch, M. Ducloy, J. Opt. Soc. Am. B 20, 793 (2003).

7. 7.G. Nikogosyan, D. Sarkisyan, Yu. Malakyan, J.Opt. Technol. 71, 602 (2004).

8. 8. B. Demtreder "Laser Spectroscopy" M., Nauka, 1985г.

9. 9. J.H. Eberly, Phys. Rev. Lett. 37, 1387 (1976).

10. 10. G.S. Agarwal, Phys. Rev. Lett. 37, 1383 (1976).

11. 11. P. Zoller, J. Phys. B: At. Mol. Phys 10, L321 (1977).

12. 12. G.S. Agarwal, Phys. Rev. A 18, 1490 (1978).

13. 13. B.J. Dalton and P.L. Knight, Optics. Commun. 42, 411 (1982).

14. 14. A.Sargsyan, D.Sarkisyan, et al., Optics and Spectroscopy, 101,762 (2006).

15. 15. A.Sargsyan, D.Sarkisyan, A.Papoyan, Phys. Rev. A 73, 033803 (2006).

16. 16. N.Papageorgiou, A.Weis, et al., Appl. Phys. B 59, 123 (1994).

17. C. Andreeva, S. Cartaleva, L. Petrov, S. M. Saltiel, D. Sarkisyan, T. Varzhapetyan, D. Bloch, M. Ducloy, Phys. Rev. A 76, 013837 (2007).

Figure Captions

Fig.1. Experimental setup. DL is the diode laser, 2 is the cell with Cs atoms, nano-cell is the cell with smoothly variable thickness of atomic vapor column in the range $L = 350 \div 5100$ nm, 3 photodetectors, $F$ are the filters.

Fig.2. Transmission spectra of the nano-cell for the transition Fg = 4 → Fe = 3 at $L= \lambda/2$, and of the cell with $L$ = 3 cm, *1* is the transmission spectrum at $L= \lambda/2$ which is well described by the « pseudoVoigt2 » function with FWHM of ~75 MHz, the arm-side temperature is 110÷120 $^0$C, *3* is the experimental Doppler broadened transmission spectrum of the cell with $L$ = 3 cm well described by the Gaussian profile *4* with FWHM of ~ 350 MHz.

Fig.3. (a) Resonant transmission spectra for the transition Fg = 4 → Fe = 3 when changing the thickness from $L = \lambda/2$ to $5\lambda$ with the step $\lambda/2$, at laser EI ~ 0.03 mW/cm$^2$ ; (b) the theoretical spectra for the same transition at $\Omega$ = 0.01 $\gamma_N$; $\gamma_N$ ~ 4.56 MHz , $\gamma_L$ ~ 6 MHz , the thermal velocity $v_T$ = 200 m/s. The lower curves in Figs. 3a, 4a, 5a have been obtained by the known SA technique with $L$= 3 cm.

Fig.4.(a) experimental resonant transmission spectra, all the parameters are the same as in Fig. 3a but laser EI ~ 0.5 mW/cm$^2$ ; (b) theoretical transmission spectra for the same transition, all the parameters are the same as in Fig. 3b but $\Omega$ = 0.06 $\gamma_N$.

Fig.5. (a) experimental resonant transmission spectra, all the parameters are the same as in Fig. 3a but laser EI ~ 7 мW/см² ; (b) theoretical transmission spectra for the same transition, all the parameters are the same as in Fig. 3b but $\Omega = 0.2\ \gamma_N$.

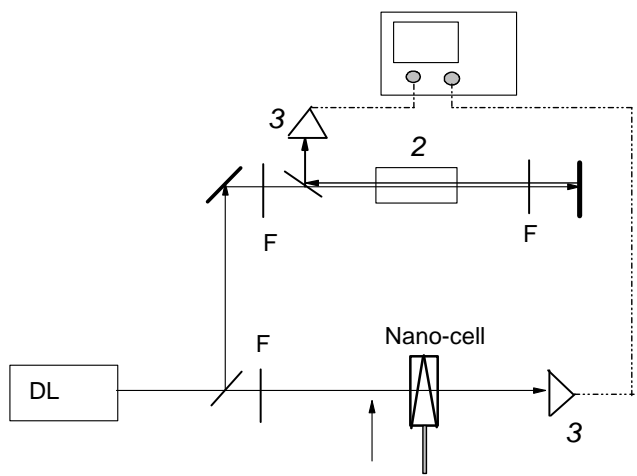

Fig. 1

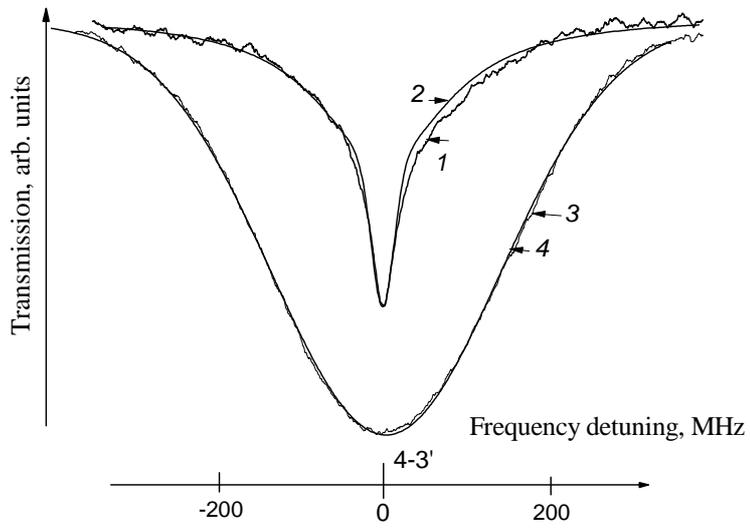

Fig. 2

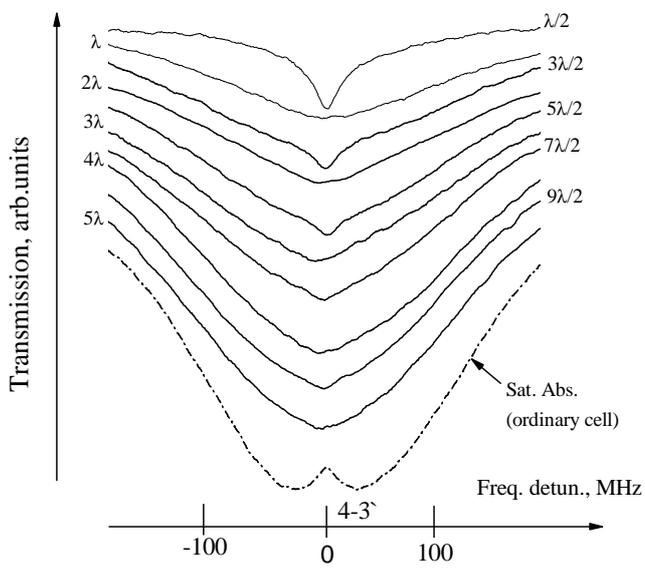

Fig. 3a

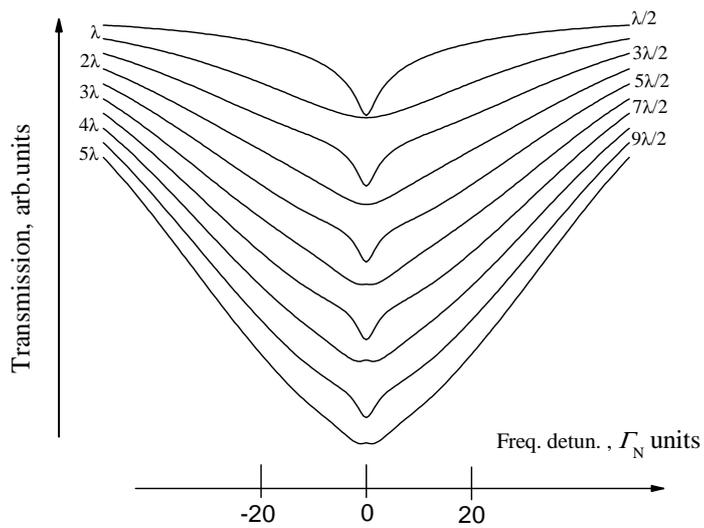

λ          λ/2
2λ          3λ/2
3λ          5λ/2
4λ          7λ/2
5λ          9λ/2

Transmission, arb.units

Freq. detun. , $\Gamma_N$ units

-20          0          20

Fig. 3b

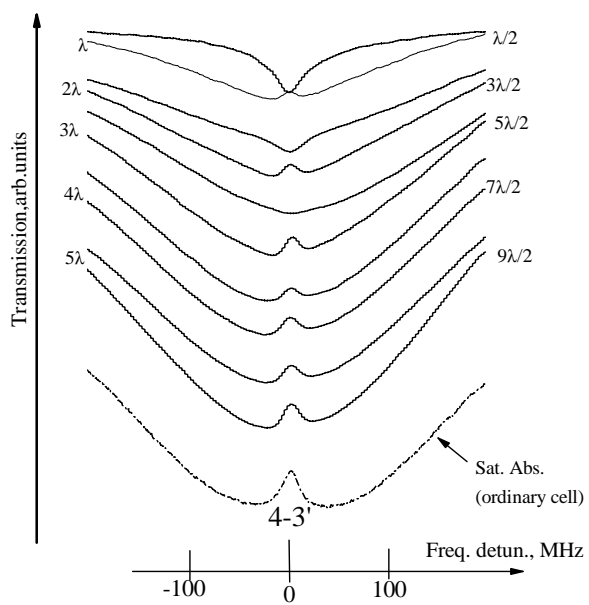

λ          λ/2
2λ          3λ/2
3λ          5λ/2
4λ          7λ/2
5λ          9λ/2

Transmission,arb.units

Sat. Abs.
(ordinary cell)

4-3'

Freq. detun., MHz

-100          0          100

Fig. 4a

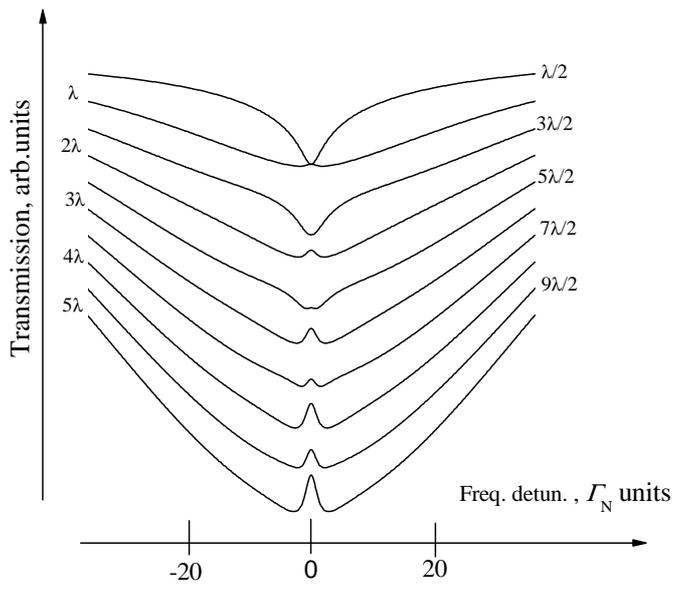

Fig. 4b

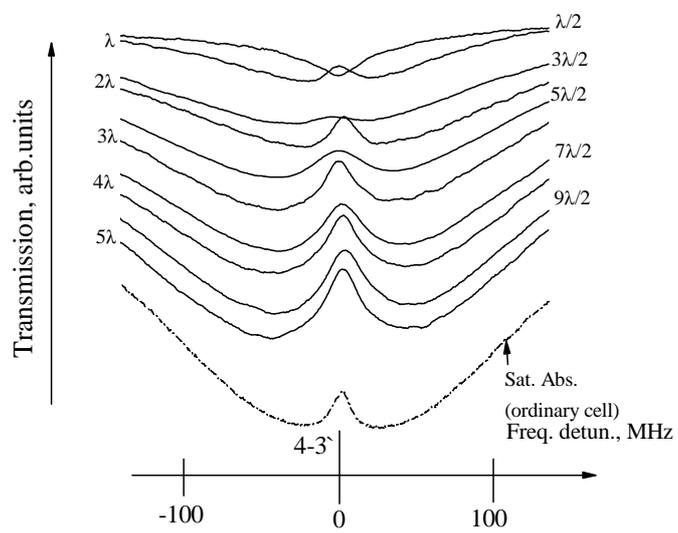

Fig. 5a

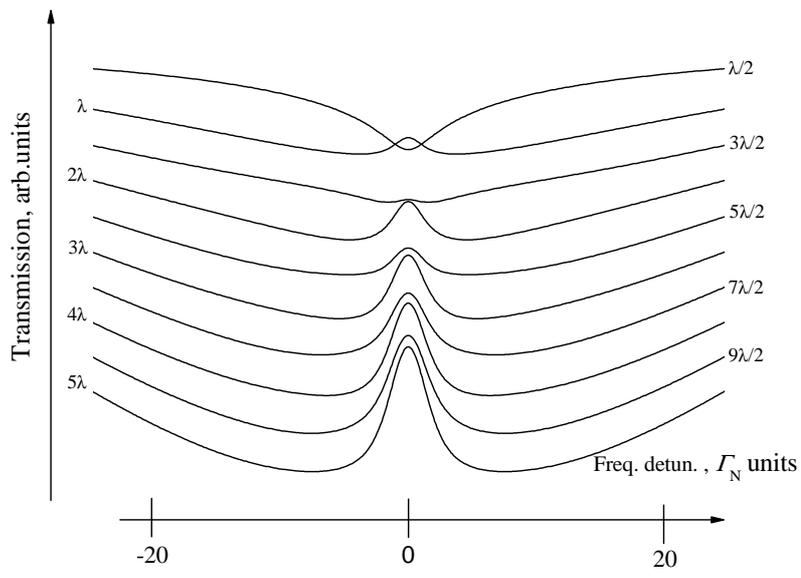

Fig. 5b